\documentstyle[aps,prl,epsf,floats]{revtex}
\begin{document}
\twocolumn[\hsize\textwidth\columnwidth\hsize\csname@twocolumnfalse%
\endcsname

\def\r{{\bf r}}

\title{Hydrogen atom in a spherical well: linear approximation}
\author{David Djajaputra and Bernard R. Cooper}
\address{
Department of Physics, West Virginia University, PO BOX 6315, Morgantown, WV 26506, USA}
\date{\today}
\maketitle
\begin{abstract}
We discuss the boundary effects on a quantum system by examining
the problem of a hydrogen atom in a spherical well. By using an 
approximation method which is linear in energy we calculate the 
boundary corrections to the ground-state energy and wave function.
We obtain the asymptotic dependence of the ground-state energy 
on the radius of the well.  
\end{abstract}
\pacs{PACS number(s): not available}]

\section{Introduction}

The problem of a hydrogen atom confined in a sphere has quite a long history in 
quantum physics. It was first investigated more than sixty years ago by Michels, 
de Boer, and Bijl\cite{michels37} in their study of the effects of pressure 
on the polarizability of atoms and molecules. This problem was then taken up
by Sommerfeld and Welker\cite{sommerfeld38} who studied the problem in detail
and calculated the critical radius for which the binding 
energy becomes zero. Over the years there has been a steady flow of papers on this
and other closely related problems.
\cite{degroot46,suryanarayana76,ludena77,weil79,leykoo79,leykoo80,lesar81,wilcox89,goldman92}
The model has often been used as a test problem
for various perturbation methods.
Using their boundary perturbation method, Hull and Julius\cite{hull56} 
obtained a formula which expresses the change of 
energy for the eigenstates in the confined system in terms of the corresponding
wave functions in free space. This method has been improved and generalized by
many authors.\cite{singh64,gorecki87} Some variational methods have also been
used to study this problem.\cite{arteca84,marin91,marin95} Fr\"oman, Yngve, and
Fr\"oman have developed the phase-integral method as a general method to attack
the problem of confined quantum systems and their 1987 paper\cite{froman87}
provides 80 references on this problem.

In recent years there has been some renewed interest on this 
problem.\cite{varshni97,varshni98} This is partly driven by the technological 
advances, such as in the field of semiconductor quantum dots,\cite{jacak98}
that have enabled the construction of interesting nanostructures which 
contain a small and controllable number (1-1000) of electrons. The computation
of the electronic structure of such systems necessarily has to take into account
the presence of the finite confining boundaries and their influence on the system.

In this paper we shall study the boundary corrections for a hydrogen atom in 
a spherical well using an approximation method which is linear in energy.
This is a well-known method in solid-state physics and has been widely used
in electronic structure calculations, under the name of Linear Muffin-Tin Orbital 
(LMTO) method,\cite{skriver84} since its initial introduction by 
O. K. Andersen in 1975.\cite{andersen,kumar} The method is best applied to the 
calculations of the wave functions of a hamiltonian with energies which 
are in close vicinity of the energy of a known wave function. To the best of our knowledge,
this simple method has never been applied previously to this problem of 
hydrogen atom in an impenetrable sphere. The present paper seeks to serve two purposes. First,
it presents a new approach, which has some pedagogical simplicity, 
to the confined hydrogen atom problem. Second, it offers an analytically tractable
problem from which one can hopefully gain some insights into the workings and
the accuracy of the LMTO method.

\section{Linear Method}

\begin{figure}
      \epsfysize=60mm
      \centerline{\epsffile{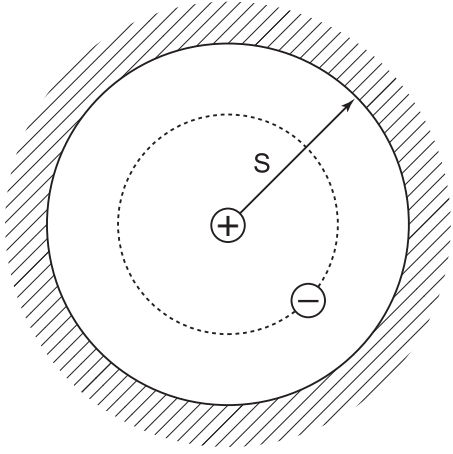}}
\smallskip
\caption{Hydrogen atom in a spherical well of radius $S$.} 
\label{spherical.well}
\end{figure}

In this paper we will examine the boundary corrections for a 
hydrogen atom situated at the center of a spherical cavity of radius $S$
as shown in Fig.\ref{spherical.well}. We will assume the wall of 
the cavity to be impenetrable and consider the following 
spherically-symmetric potential:

\begin{equation}
V(r) = \cases{ -e^2/r, & $r < S,$ \cr
\infty, & $r > S.$}  
\end{equation} 

\noindent The radius of the cavity will be assumed 
to be much larger than the Bohr radius: $S \gg a_0.$ 
In the remainder of the paper we shall use the atomic units:

\begin{equation}
\hbar = {e^2 \over 2} = 2m = 1.
\end{equation}

\noindent The unit of length is the Bohr radius
$a_0 = \hbar^2/me^2$ and the unit of energy is the Rydberg:
${\rm Ry} = e^2/2a_0 = 13.6$ eV. The Schr\"odinger equation takes the following
form:

\begin{equation}
H \Psi( {\bf r} ) = \Big( - \nabla^2 - {2 \over r} \Big) \Psi( {\bf r} ) = 
E \Psi( {\bf r} ).
\label{schrodinger}
\end{equation}

\noindent The wave function $\Psi({\bf r})$ satisfies the Schr\"odinger equation 
for the hydrogen atom for $r < S$, in particular it should still be regular at the origin.
The only difference from the free-space case is that now we have to impose a different 
boundary condition: the wave function should vanish at $r = S$ instead of
at $r = \infty$.

For $S \gg a_0$, the changes in the ground-state wave function and energy 
due to the presence of the wall are expected to be ``small'' because the 
wave function is concentrated at the center of the cavity, far away from the
confining wall. 

In free space, i.e. in the absence of the confining
cavity, the hydrogen atom has the familiar Rydberg spectrum:

\begin{equation}
\varepsilon_n = - {1 \over n^2}, \quad n = 1,2, \dots \ . 
\label{spectrum}
\end{equation}

\noindent In the presence of the cavity, we write

\begin{equation}
E_n = \varepsilon_n + \Delta \varepsilon_n.
\end{equation}

\noindent We use small letters ($\varepsilon, \psi$, etc.) to denote quantities 
for the free-space problem and capital letters ($E, \Psi$, etc.) for
the corresponding quantities in the cavity problem. The dimensionless parameter
$(\Delta \varepsilon_n / \varepsilon_n)$ is expected to be small for 
$n^2 a_0 \ll S.$ In the linear method, the (unnormalized) wave function at 
energy $E_n$ is approximated by

\begin{equation}
\Psi(E_n, {\bf r}) = \psi(\varepsilon_n, {\bf r}) + \Delta \varepsilon_n \
\dot{\psi}(\varepsilon_n, {\bf r}). 
\end{equation}

\noindent Here $\dot{\psi}(\varepsilon_n, {\bf r})$ is the derivative with 
respect to energy of $\psi(\varepsilon, {\bf r})$ evaluated at $\varepsilon
= \varepsilon_n$:

\begin{equation}
\dot{\psi}(\varepsilon_n, {\bf r}) = [ \partial \psi(\varepsilon, {\bf r})
/ \partial \varepsilon ] \ (\varepsilon = \varepsilon_n).
\end{equation}

\noindent The eigenfunctions in the cavity problem are then obtained by
imposing the boundary condition at $r = S$:

\begin{equation}
\Psi(E_n, S,\hat{{\bf r}}) = 0,
\end{equation}

\noindent which gives an expression for the energy correction:

\begin{equation}
\Delta \varepsilon_n = - {\psi(\varepsilon_n, S,\hat{{\bf r}})
\over \dot{\psi}(\varepsilon_n, S,\hat{{\bf r}})}.
\label{energycorrection}
\end{equation}

\noindent Here $\hat{{\bf r}}=(\theta,\phi)$ is a unit vector 
in the direction of ${\bf r}.$

To apply the linear approximation method we need the general solution to
the Schr\"odinger equation at an arbitrary energy $E$. Since we are
dealing with a spherically-symmetric system, we can separate the
variables:

\begin{equation}
\Psi ({\bf r}) = R(r) Y_{lm} (\hat{{\bf r}}).
\end{equation}

\noindent The resulting radial differential equation is

\begin{equation}
{d^2 R \over dr^2} + {2 \over r} {dR \over dr} + 
\Big[ E + {2 \over r} - {l(l+1) \over r^2} \Big] R = 0.
\end{equation}

\noindent Transforming the variables by defining

\begin{equation}
\omega = \sqrt{-E}, \qquad \rho = 2 \omega r,
\end{equation}

\noindent and using the following trial functional form\cite{seaborn}

\begin{equation}
R(\rho) = \rho^l e^{-\rho / 2} u(\rho),
\end{equation}

\noindent then gives us the following differential equation\cite{seaborn}

\begin{equation}
\rho u'' + \Big[ 2(l+1) - \rho \Big] u' 
- \Big[ l+1 - {1 \over \omega} \Big] u = 0, 
\end{equation}

\noindent which is the equation for the confluent hypergeometric function. 
The general solution of this equation, which is regular at the origin, is\cite{seaborn}

\begin{equation}
u(\rho) = A \ {}_1F_1\Big(l+1-{1 \over \omega}; 2l+2; \rho \Big), 
\end{equation}

\noindent where $A$ is a normalization constant. The radial part of the 
general solution to the Schr\"odinger equation
Eq.(\ref{schrodinger}) with energy $E = - \omega^2$ therefore is

\begin{equation}
R_l(\omega,r) = A \ (2 \omega r)^l e^{-\omega r} 
{}_1F_1\Big(l+1-{1 \over \omega}; 2l+2; 2 \omega r \Big).
\label{radialfunction}
\end{equation}

\noindent The free-space solution is obtained by requiring that $R(r) \rightarrow 0$
as $r \rightarrow \infty.$ From the properties of the hypergeometric functions,\cite{seaborn} 
this can only happen if $(l+1- 1/\omega)$ is a negative integer or zero. This implies that

\begin{equation}
{1 \over \omega} = n, \qquad l = 0,1, \dots, (n-1),
\end{equation}

\hyphenation{Ryd-berg}

\noindent with $n$ a positive integer. This directly leads to the Rydberg
spectrum in Eq.(\ref{spectrum}).

\begin{figure}
      \epsfysize=50mm
      \centerline{\epsffile{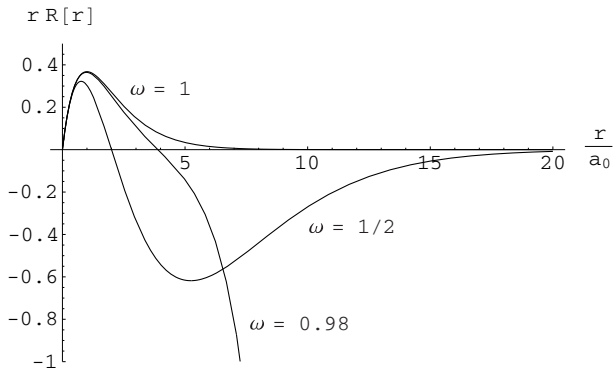}}
\smallskip
\caption{The function $r R_l(\omega,r)$ as a function of $r/a_0$ for $l=0$
and $\omega$ = 1, 0.98, and 0.50. The $\omega = 1$ curve is nodeless. As $\omega$
is decreased from 1 to 0.50, the node of the wave function moves from $r = \infty$
to $r = 2 a_0$.}
\label{wave.function}
\end{figure}

The function $R_l(\omega,r)$ is plotted in Fig.\ref{wave.function} for $l=0$
and $\omega$ = 1, 0.98, and 0.50. The $\omega = 1$ curve is the ground-state 
wave function of the hydrogen atom in free space and is nodeless.
Here a node of $R_l(\omega,r)$ is defined to be a value of the argument $r$ 
which gives zero value for the function $R_l(\omega,r)$. As $\omega$
is reduced below 1, the wave function acquires a single node which moves from $r = \infty$
to $r = 2a_0$ at $\omega = 0.50.$ where it becomes the $(n,l)=(2,0)$ eigenstate of 
the hydrogen atom in free space. One therefore can obtain the ground-state wave
function and energy of the hydrogen atom in a cavity of radius $S$ by numerically
searching for the energy which gives a wave function with a single node at $r = S$. 
This provides a useful comparison for our approximation.

Since the spherical harmonics are independent of the energy we can recast
Eq.(\ref{energycorrection}) into 

\begin{equation}
\Delta \varepsilon_{nl} = 2 \omega_n {R_l(\omega_n,S) \over \dot{R}_l(\omega_n,S)}.
\label{radialratio}
\end{equation}

\noindent where $\omega_n = \sqrt{-\varepsilon_n}$ and 

\begin{equation}
\dot{R}_l(\omega_n,S) = [\partial R_l(\omega,S)/\partial \omega] \ (\omega = \omega_n). 
\label{dotradial}
\end{equation}

\noindent Substituting the radial function $R_l(\omega,r)$ in Eq.(\ref{radialfunction})
into Eq.(\ref{radialratio}) then gives us an explicit formal expression 
for $\Delta \varepsilon_n$ which should be valid for $R \gg n^2 a_0.$
Note that the presence of the finite boundary lifts the azimuthal degeneracy of the states 
with different orbital quantum number $l$ (and the same radial 
quantum number $n$).\cite{shea96}
As in the case of the screened Coulomb potential, this occurs because one no
longer deal with the pure Coulomb potential.\cite{greiner,symmetry} 

To gain an insight into Eqs.(\ref{radialratio})-(\ref{dotradial}), 
we shall consider the ground state
($n = 1$), which is a special case of the zero angular momentum ($l = 0$) states.
We have

\begin{equation}
R_0(\omega,r) = A \ e^{-\omega r} 
{}_1F_1\Big(1-{1 \over \omega}; 2; 2 \omega r \Big).
\label{angularzero}
\end{equation}

\noindent For the ground state ($n=1$), this is

\begin{equation}
R_0(1,r) = A \ e^{-r} 
{}_1F_1\Big(0; 2; 2 r \Big) = A \ e^{-r}.
\label{groundstate}
\end{equation}

\noindent We are interested in obtaining a simple analytical expression 
of the correction to the ground-state energy for $S \gg a_0$, therefore we need to calculate 
the limiting form of $\dot{R}_0(\omega,r)$ for $r \gg a_0.$ The asymptotic expansion
of the hypergeometric function ${}_1F_1(a,b,z)$ for large $z$ is\cite{abramowitz}

\begin{equation}
{ {}_1F_1(a,b,z) \over \Gamma(b) } = {e^{i \pi a } \over z^a} {I_1(a,b,z) \over
\Gamma(b-a)} + e^z z^{a-b} {I_2(a,b,z) \over \Gamma(a)},
\label{asymptotic}
\end{equation}

\noindent with

\begin{equation}
I_1(a,b,z) = \sum_{n=0}^{R-1} {(a)_n (1 + a - b)_n \over n!}
{e^{i \pi n} \over z^n} + {\cal O}( |z|^{-R} ),
\end{equation}

\begin{equation}
I_2(a,b,z) = \sum_{n=0}^{R-1} {(b-a)_n (1-a)_n \over n!} {1 \over z^n}
+ {\cal O}( |z|^{-R} ). 
\end{equation}

\noindent The Pochhammer symbol $(a)_n$ is defined by\cite{seaborn}

\begin{equation}
(a)_n = a (a+1) \cdots (a + n - 1) = {\Gamma(a + n) \over \Gamma(a)}. 
\end{equation}

\begin{figure}
      \epsfysize=80mm
      \centerline{\epsffile{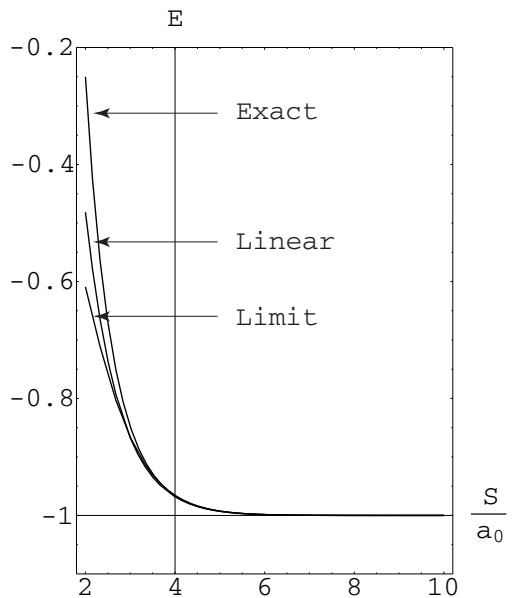}}
\smallskip
\caption{Dependence of the ground-state energy of a hydrogen atom 
confined in a spherical cavity on the radius of the cavity $S$. 
The topmost curve is the exact result which is obtained by numerically 
searching for the node of the wave function.The middle 
curve is obtained from the linear approximation, Eq.(\ref{radialratio}), 
using the exact wave function Eq.(\ref{groundstate}). The lowest curve 
is obtained using the limiting formula Eq.(\ref{limitcorrection}). } 
\label{accuracy}
\end{figure}

\noindent We need to calculate the derivative of this function at 
$a = (1 - 1/\omega)$ with $\omega = 1.$ In this case the dominant term comes from 
the derivative of $\Gamma(a)$ in the second term in Eq.(\ref{asymptotic}).
The first term can be neglected because it does not have the exponential
term $e^z$ which dominates the derivative at large distances. Keeping only
the largest term, we get

\begin{equation}
{\partial \over \partial a} {}_1F_1(a,b,z) \approx - e^z z^{a-b} \ \Gamma(b) 
I_2(a,b,z) \ {\psi(a) \over \Gamma(a)}. 
\end{equation}

\noindent Here $\psi(a)$ is the digamma function: 
$\psi(a) = \Gamma'(a)/\Gamma(a)$.\cite{abramowitz}
Its ratio with $\Gamma(a)$ as $a \rightarrow 0$ is

\begin{equation}
\lim_{a \rightarrow 0} {\psi(a) \over \Gamma(a)} =
\lim_{a \rightarrow 0} \ {- \gamma - 1/a \over - \gamma + 1/a} = -1, 
\end{equation}

\noindent where $\gamma$ is the Euler constant. This then gives

\begin{equation}
\Big[ {\partial \over \partial a} {}_1F_1(a,b,z) \Big] (a \rightarrow 0) 
\approx e^z z^{a-b} \ \Gamma(b) I_2(a,b,z). 
\end{equation}

\noindent Using this expression, and keeping only the first two terms in 
$I_2(a,b,z)$, we can obtain the limiting form of $\dot{R}_0(\omega,r)$
at large $r$ and $\omega \rightarrow 1$:

\begin{equation}
\dot{R}_0(\omega,r) \approx {A e^{-\omega r} \over \omega^2}
\Big\{ { e^{2 \omega r} \over (2 \omega r)^{1 + 1/\omega} }
\Big[ 1 + { \Gamma (2 + 1/ \omega) \over 2 \omega r \ 
\Gamma(1 / \omega) } \Big] \Big\}.
\end{equation}

\noindent Exactly at $\omega = 1$, this expression becomes

\begin{equation}
\dot{R}_0(1,r) \approx {A e^{r} \over 4 r^2}
\Big[ 1 + {1 \over r} \Big].
\end{equation}

\noindent Finally, using this equation and Eq.(\ref{groundstate}) in
Eq.(\ref{radialratio}), we get the boundary correction to the ground-state 
energy:

\begin{equation}
\Delta \varepsilon_0 (S) \approx 8 S(S-1) e^{-2S}, \quad
S \gg a_0.
\label{limitcorrection}
\end{equation}

\section{Discussion}

Fig.\ref{accuracy} displays the asymptotic dependence of the
energy correction on the radius of cavity, Eq.(\ref{limitcorrection}),
together with the exact curve and
the curve obtained from the linear approximation method, Eq.(\ref{radialratio}), using
the exact wave function Eq.(\ref{groundstate}). It is seen that the asymptotic formula, 
Eq.(\ref{limitcorrection}), is fairly accurate for radii greater than
about four Bohr radius. Note that the exact energy at $S = 2 a_0$ is 
equal to ${1 \over 4}$ Ry, which is the energy of the first excited 
state $(n,l)=(2,0)$ of the hydrogen atom in free space. This is because
the corresponding wave function has a node at $r = 2 a_0$ as can be seen
in Fig.\ref{wave.function}. 

The asymptotic formula Eq.(\ref{limitcorrection}), which is the 
limit curve in Fig.\ref{accuracy}, is a ``double-approximation'' to the exact curve. 
It is an asymptotic form of the linear curve, Eq.(\ref{radialratio}), 
valid for large values of $S/a_0$. The linear curve itself is an approximation, 
linear in energy, to the exact curve. For small values of $S/a_0$, and within
the linear approximation method, one has to use Eq.(\ref{radialratio}) which
in general, unfortunately, does not correspond to a simple analytic expression. 
This does not pose a problem in actual electronic-structure applications 
because there the wave function and its energy derivative are computed numerically. 
In this paper, for pedagogical purposes, we have calculated the asymptotic
formula, Eq.(\ref{limitcorrection}), which does correspond to a simple 
analytic expression.

Knowing the dependence of the ground-state energy on the cavity radius, 
Eq.(\ref{limitcorrection}), allows us to calculate the pressure needed to 
``compress'' a hydrogen atom in its ground state to a certain size. This is
given by

\begin{equation}
p(S) = - {\partial \Delta \varepsilon_0 \over \partial V} 
\approx {4 e^{-2 S} \over \pi} \Big( 1 - {2 \over S} \Big).
\end{equation} 

\noindent At $S = 4 a_0$ this has a value of $2.13 \times 10^{-4}$ eV/$a_0^3$
= $1.47 \times 10^4$ GPa. At this radius, the change of the ground-state energy
is 0.032 Ry which is only three percent of the binding energy of a free hydrogen
atom.

The information on the effects of the boundary on the wave function of the atom can
also be used to study the influence of the boundary on other properties of the atom, 
e.g., the spin-orbit coupling energy. It is also interesting to calculate the changes
in the wave function and energy of the atom when it is displaced from the center of 
the cavity, and the force that will push it back towards the center. The linear method
also seems to be well-suited for the analysis of the ``soft-cavity'' case where we have 
a finite potential outside the cavity, instead of the infinite potential 
considered in this paper. These topics will be examined in future works.

In conclusion, we have used a linear approximation method to calculate the
asymptotic dependence of the ground-state energy of a hydrogen atom confined
to a spherical cavity on the radius of the cavity. The boundary correction to
the energies of the excited states can be obtained using the same method. 

\bigskip

\noindent {\it Acknowledgements}---D. D. is grateful to Prof. David L. 
Price (U. Memphis) for introducing him to Andersen's linear approximation 
method and for many useful discussions. Thanks are also due to Dr. H. E. Montgomery, Jr.
for many useful references. This work has been supported by 
AF-OSR Grant F49620-99-1-0274.


\begin{thebibliography}{99}

\bibitem{michels37} A. Michels, J. de Boer, and A. Bijl, ``Remarks concerning
molecular interaction and their influence on the polarisability,'' Physica {\bf 4},
981--994 (1937).

\bibitem{sommerfeld38} A. Sommerfeld and H. Welker, ``K\"unstliche Grenzbedingungen
beim Keplerproblem,'' Ann. Phys. {\bf 32}, 56--65 (1938).

\bibitem{degroot46} S. R. de Groot and C. A. ten Seldam, ``On the energy levels of
a model of the compressed hydrogen atom,'' Physica {\bf 12}, 669--682 (1946).

\bibitem{suryanarayana76} D. Suryanarayana and J. A. Weil, ``On the hyperfine
splitting of the hydrogen atom in a spherical box,'' J. Chem. Phys. {\bf 64},
510--513 (1976).

\bibitem{ludena77} E. V. Ludena, ``SCF calculations for hydrogen in a spherical
box,'' J. Chem. Phys. {\bf 66}, 468--470 (1977).

\bibitem{weil79} J. A. Weil, ``Hydrogen atom in a spherical box. II. Effect on 
hyperfine energy of excited state admixture,'' J. Chem. Phys. {\bf 71}, 2803--2805 (1979).

\bibitem{leykoo79} E. Ley-Koo and S. Rubinstein, ``The hydrogen atom within spherical
boxes with penetrable walls,'' J. Chem. Phys. {\bf 71}, 351--357 (1979).

\bibitem{leykoo80} E. Ley-Koo and S. Rubinstein, ``The hydrogen atom inside boxes
with paraboloidal surfaces,'' J. Chem. Phys. {\bf 73}, 887--893 (1980).

\bibitem{lesar81} R. LeSar and D. R. Herschbach, ``Electronic and vibrational
properties of molecules at high pressures. Hydrogen molecule in a rigid spherical box,''
J. Phys. Chem. {\bf 85}, 2798--2804 (1981).

\bibitem{wilcox89} W. Wilcox, ``A formula for energy displacements for the 
confined hydrogen atom,'' Am. J. Phys. {\bf 57}, 526--528 (1989).

\bibitem{goldman92} S. Goldman and C. Joslin, ``Spectroscopic properties of 
an isotropically compressed hydrogen atom,'' J. Phys. Chem. {\bf 96}, 6021--6027 (1992).

\bibitem{hull56} T. E. Hull and R. S. Julius, ``Enclosed quantum mechanical systems,''
Can. J. Phys. {\bf 34}, 914--919 (1956).

\bibitem{singh64} K. K. Singh, ``Theory of boundary perturbation and the 
compressed hydrogen molecular ion,'' Physica {\bf 30}, 211--222 (1964).

\bibitem{gorecki87} J. Gorecki and W. B. Brown, ``Iterative boundary perturbation
method for enclosed one-dimensional quantum systems,'' J. Phys. B {\bf 20}, 5953--5957
(1987).

\bibitem{arteca84} G. A. Arteca, F. M. Fernandez, and E. A. Castro, ``Approximate
calculation of physical properties of enclosed central field quantum systems,''
J. Chem. Phys. {\bf 80}, 1569--1575 (1984).

\bibitem{marin91} J. L. Marin, S. A. Cruz, ``On the use of direct variational 
methods to study confined quantum systems,'' Am. J. Phys. {\bf 59}, 931--935 (1991).

\bibitem{marin95} J. L. Marin, R. Rosas, and A. Uribe, ``Analysis of asymmetric 
confined quantum systems by the direct variational method,'' Am. J. Phys.
{\bf 63}, 460--463 (1995).

\bibitem{froman87} P. O. Fr\"oman, S. Yngve, and N. Fr\"oman, ``The energy levels and 
the corresponding normalized wave functions for a model of a compressed atom,'' 
J. Math. Phys. {\bf 28}, 1813--1826 (1987).

\bibitem{varshni97} Y. P. Varshni, ``Accurate wavefunctions for the confined hydrogen
atom at high pressures,'' J. Phys. B {\bf 30}, L589--L593 (1997), and references therein.

\bibitem{varshni98} Y. P. Varshni, ``Critical cage radii for a confined hydrogen atom,''
J. Phys. B {\bf 31}, 2849--2856 (1998).

\bibitem{jacak98} L. Jacak, P. Hawrylak, and A. W\'ojs, {\it Quantum Dots},
Springer-Verlag, Berlin, 1998.

\bibitem{skriver84} H. L. Skriver, {\it The LMTO Method}, Springer-Verlag, Berlin, 1984.

\bibitem{andersen} O. K. Andersen, ``Linear methods in band theory,'' 
Phys. Rev. B{\bf 12}, 3060--3083 (1975).

\bibitem{kumar} V. Kumar, O. K. Andersen, and A. Mookerjee, {\it 
Lectures on Methods of Electronic Structure Calculations}, World Scientific,
Singapore, 1994.

\bibitem{seaborn} J. B. Seaborn, {\it Hypergeometric Functions and Their
Applications}, Springer-Verlag, New York, 1991, Chapter 6.

\bibitem{shea96} R. W. Shea and P. K. Aravind, ``Degeneracies of the spherical
well, harmonic oscillator and hydrogen atom in arbitrary dimensions,''
Am. J. Phys. {\bf 64}, 430--434 (1996).

\bibitem{greiner} W. Greiner and B. M\"uller, {\it Quantum Mechanics:
Symmetries}, Springer-Verlag, Berlin, 1994, Chapter 14.

\bibitem{symmetry} In group theoretical language,
modifications to the pure Coulomb potential break the SO(4) symmetry 
of the hydrogen atom: the Runge-Lenz {\it operator} no longer commute with the
hamiltonian. This should be contrasted with the classical
case where the Runge-Lenz {\it vector} is still a good constant of motion 
and the presence of the boundary does not have any effect on 
the orbit of the particle if it is greater than the orbit's aphelion.
Greiner's book\cite{greiner} gives a detailed discussion on the Runge-Lenz vector and 
the SO(4) symmetry of the hydrogen atom.

\bibitem{abramowitz} M. Abramowitz and I. A. Stegun, {\it Handbook of 
Mathematical Functions}, Dover, New York, 1965, Eq.(13.5.1).

\end{thebibliography}
\end{document}